\documentclass{elsart}
\usepackage{graphicx}
\usepackage{amssymb}
\begin{document}
\begin{frontmatter}

\title
{Directed Accelerated Growth:  Application in  Citation Network
}
\author
{
	Parongama Sen
}
\address
{
	Department of Physics, University of Calcutta,
	    92 Acharya Prafulla Chandra Road, Kolkata 700009, India.
}

\begin{abstract}

In many real world networks, the number of links increases
nonlinearly with the number of nodes. Models of such accelerated 
growth have been considered earlier with deterministic and stochastic
number of links. Here we consider
stochastic accelerated growth in a network where links are directed.
With the number of out-going links following a  power law distribution,
the results are similar to the 
undirected case. As the accelerated growth is enhanced, the degree
distribution becomes independent of the  ``initial attractiveness'', 
a parameter which plays a key role
in directed networks.
As an example of a directed model with accelerated growth, the citation network is considered,
in which  the  distribution  of the number of outgoing link has an exponential tail.
 The role of accelerated growth is examined here with
 two different growth laws.

\end{abstract}
\end{frontmatter}

The number of links shows a nonlinear growth in time 
in many real-world networks which evolve in time
\cite{doro_review,ba_review}.  Examples of such network are  the  Internet
\cite{Faloutsos}, World Wide Web (WWW)  \cite{broder},  collaboration \cite{collab}, 
 word web \cite{doro_word},
citation  \cite{original,redner,citation},  metabolism  \cite{metabolism},
gene regulatory network \cite{expt1,gagen1,gagen2} etc.
The number of
links may increase in  two  ways: new nodes may tend to get attached to 
more nodes as the
 size of the network increases,
and  there may be additional links generated between the older
nodes in a non-linear fashion.
 These two factors maybe present  
either singly  or simultaneously resulting in the 
accelerated growth. In some networks 
like the citation and the gene regulatory network, new links  between older nodes
are forbidden and therefore only the  first scheme is valid,  
while in collaboration
network or internet, the second factor is dominating.

In general, in a network with  accelerated growth, the total number of
links $l $ shows a non-linear increase in time, i.e., $l(t) \sim t^{\delta}$ 
with $\delta > 1$. 
Such an acceleration will be possible  
when the number of links $m(\tau)$ at any time $\tau$ 
available to the network
is not a constant but increases in such a manner that
$l(t) = \Sigma_{\tau=1}^t  m(\tau)$  
shows a non-linear increase with time.

The case  when the new node gets a fixed  number of links but
older nodes get new links the number of which grows in a non-linear 
fashion has been 
considered in both isotropic and directed models of growing networks 
\cite{collab,doro_word,doro1}, 
showing that
it is distinct from networks with a linear growth rule. 
Networks in which older nodes do not get new links have also
been considered recently, with the possibility of  a new node to have either
 deterministic or stochastic number of links \cite{ps}.

Many real systems like
the citation network and WWW 
 are directed
networks \cite{citation,tadic}. In the citation network, accelerated growth has been shown
to take place \cite{citation}. This network also evolves in such a way that new 
links are forbidden between old nodes.  We propose in this paper directed models of accelerated growth
with number of outgoing links given by a particular distribution and where older nodes
are not allowed to get linked. One of these  models  mimics the citation network.

{\it {Directed accelerated growth with  power law distribution}}

As in \cite{ps}, here we assume that in the growing network, the number of outgoing links $m$ is determined
by a power-law distribution, i.e.,
\begin{equation}
P(m) \sim m^{-\lambda}
\end{equation}
with $1 \leq m \leq N(\tau)$
at any time $\tau$ when the number of nodes in the network is $N(\tau)$. 
The number of nodes is usually a linear function of time and 
$N(\tau) = \tau$ when nodes
are added one by one to the network.
$\lambda$ is the parameter 
controlling the measure of accelerated growth.
As a new node comes in, it will get preferentially attached to $m$ existing nodes
with the probability
\begin{equation}
\Pi  \propto A + k_{in}
\end{equation}
where $k_{in}$ is the in-degree of the existing node and $A$ is the initial
attractiveness of the network. $A$ is a parameter 
which is necessary in all directed networks to  ensure that a node even 
with $k_{in}=0$
has the chance to get linked with the new node. Without acceleration, when 
$m$ is fixed, one gets
a scale-free network with the the degree distribution $P(k) \sim k^{-\gamma}$ 
where \cite{doro2}
\begin{equation}
\gamma= 2+A/m. 
\end{equation}
With $m$ varying (growing) in time, the nature of the degree distribution $P(k)$ 
is expected to change. This is obvious if one considers the 
average value of $m$ which shows the following behaviour:

$$ \langle m\rangle = {\rm {~~const~~ for~~ }} \lambda > 2$$ 
$$ ~~~~~~~~~~~~~~~~ \propto N^{2-\lambda} {\rm~~for~~} 1 < \lambda < 2$$ 
$$ ~~~~~~~~~~~~~ \propto N {\rm ~~for~~}  \lambda < 1.$$ 

For large values of $\lambda$, one should expect that the 
 the distribution retains 
the features of a non-accelerated growing network. 
This is verified in the simulation.
In Fig. 1, the data for $P(k)$ for different $\lambda$ values are shown.
With  $A = 0.1$, $\gamma$ should be close to 2 if equation (3)
is still valid.  
The data indicates that the exponent is indeed close
to 2 for $\lambda \geq  2$.
However, for smaller values of $\lambda$, $P(k)$  is  different from that
of the non-accelarated case and not a power law at all.
$P(k)$ is a stretched exponential function for $2 \geq \lambda \geq 1$. At $\lambda=1$, $P(k)$
 decays weakly  with $k$ and in fact shows a  weak growth when $\lambda$ is further decreased
 (curves (c) and (d) in Fig. 2).
 These results are very similar to the undirected model \cite{ps}.
 The fact that for $\lambda < 2$, $P(k)$ does not have  a power-law
 decay shows that the randomness in $m$ is relevant here.

\begin{center}
\begin{figure}
\noindent \includegraphics[clip, width=8cm, angle=270]{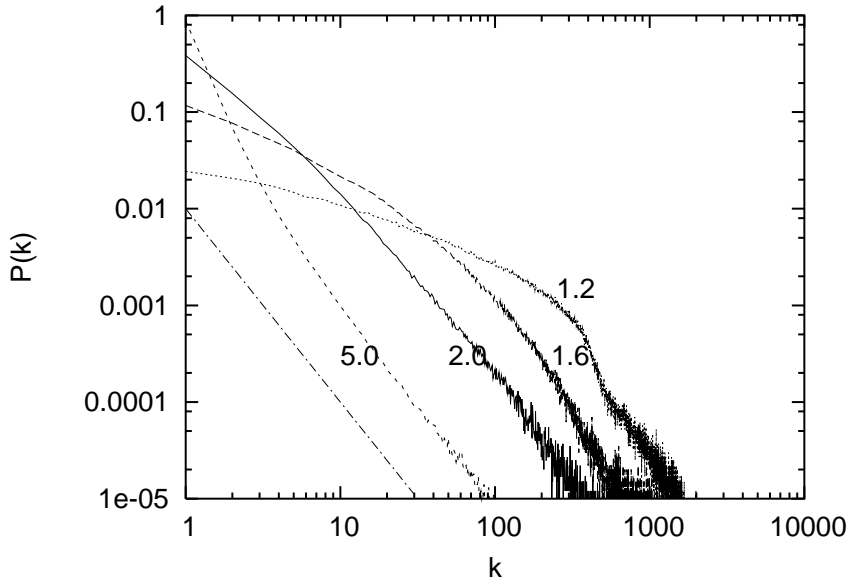}
 \caption{The total degree distribution $P(k)$ for different values of 
 $\lambda$ (shown as labels of  the curves) are plotted against $k$.
 All the data are for networks evolved upto 4000 nodes  in this and  in the
 other figures. The value of $A=0.1$.
   The straightline has slope 2.0.}
\end{figure}
\end{center}

\begin{center}
\begin{figure}
\noindent \includegraphics[clip, width=8cm, angle=270]{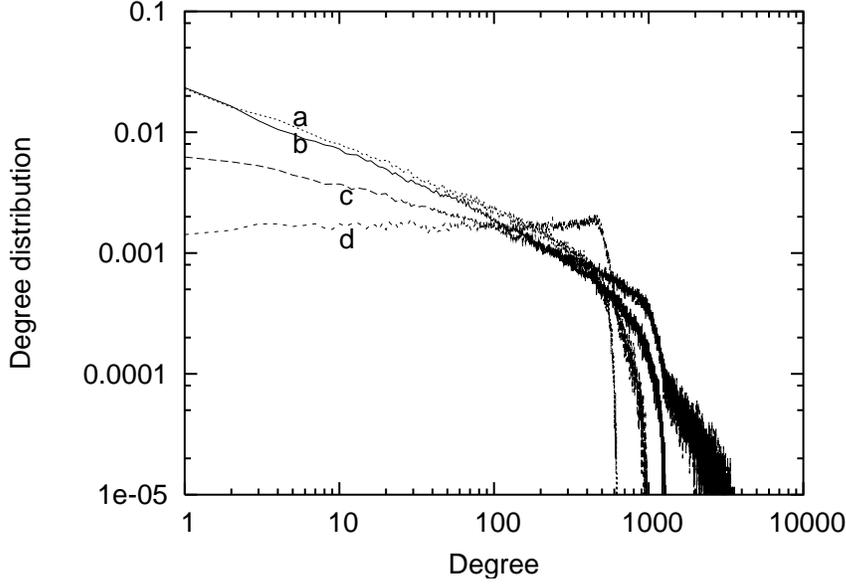}
 \caption{The in-degree distributions are shown for
 (a) $\lambda=0.6$ 
and (b) $\lambda=1.0$. The total degree distributions 
are markedly different: (c) is for $\lambda =1 $ and (d) for $\lambda=0.6.$
$A=0.1$ in this figure.
}
\end{figure}
\end{center}

From the simulations, one can directly obtain the in-degree distribution, 
which is the 
distribution of the number of incoming links to a node 
\cite{krapetal}.
 In Fig. 2, we have  plotted the in-degree distribution alongwith the
 total degree distribution as they show noticeable  difference when $\lambda$ is
 less than 2.
 To explain the above feature, we note that 
 if $P_{in} $ and $P_{out}$ are the in-degree and out-degree distributions respectively
 then 
  the total degree distribution can be expressed as
 \begin{equation}
 P(k) = \Sigma_{k_1=1}^{k_1^{max}}P_{out}(k_1)P_{in}(k-k_1)
 \end{equation}
where $k_1^{max}$ is the maximum value of the out-degree $k_1$. Evidently, if
 $k_1^{max} << k$, the in-degree and total degree distributions 
 have similar behaviour  with $k$. 
 As $\lambda$ becomes smaller, $k_1^{max}$ increases and this inequality
 does not hold good anymore and therefore the in-degree and total degree distributions
 become different.

The variation of the degree distribution with a fixed value of $\lambda$ and different $A$
values is also worth studying. In Fig. 3 we show this variation. 
The data shows that the dependence on $A$ gradually vanishes when the accelerated growth
is increased. This can be explained by noting that when $\lambda$ is small,
each incoming node has to get linked to a large number of existing nodes and therefore the
initial attractiveness factor in the attachment
probability becomes unimportant. However, it has to be  different from zero for
reasons mentioned earlier.

\begin{center}
\begin{figure}
\noindent \includegraphics[clip, width=8cm, angle=270]{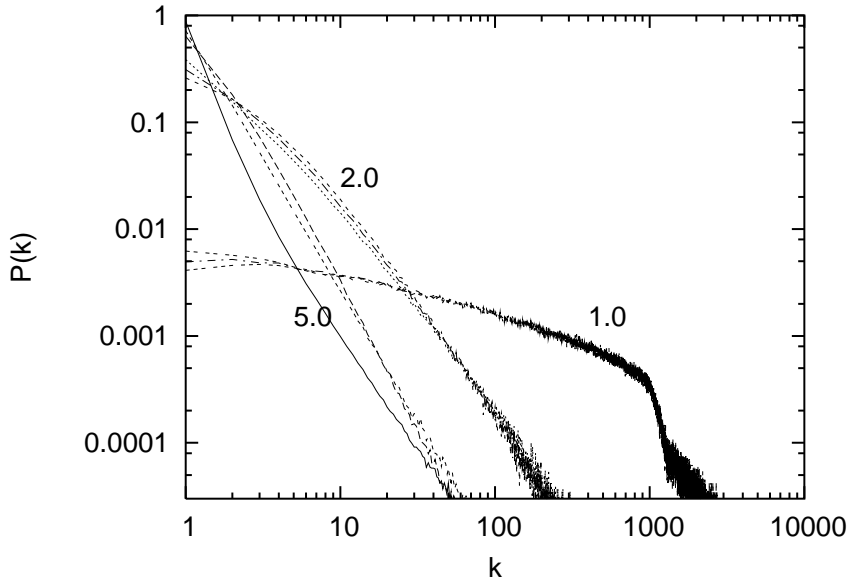}
\caption {The variation of $P(k)$ against $k$ with 
$A=0.1,0.5$ and 1 are shown for 3 fixed values of $\lambda$: 
$\lambda=5$,  $\lambda=2$ and $\lambda=1$. The dependence on $A$ is seen to
disappear as $\lambda$ becomes smaller.}
\end{figure}
\end{center}
\pagebreak

{\it {Accelerated growth model for citation network}}

The citation network is  a directed network in which the nodes are published papers.
A new paper establishes a directed
link with an old paper by citing it. Obviously new links  cannot be formed between old
nodes.

In a citation
network, one is  interested  to  find out the in-degree distribution
$P (k)$,    the number of
papers which have been cited  $k$ times.
The citation data $P(k)$ may be  studied in various ways:
 (journal-wise,
  time-wise \cite{x}) etc.
  A citation network is obviously a growing network but
  while the  in-degree distribution maybe scale-free,
  the out-degree distribution is not, as the number of papers cited is usually limited.
  However, the  average number of citations does show an increase
  with the number of published papers, indicating that it has an accelerated growth \cite{citation}.
  In the present model,
  we find out the distribution $P(k)$ irrespective of all
  details -  $P(k) $ is obtained for the set of all published papers
  and all citations made to them after their publication to date.
  As such it is difficult to compare with  the available data for citation, and instead of attempting to do so,
  we investigate the role of accelerated growth in a citation network.

  In \cite{citation}, it was shown that the average number of citations 
  (out-degree) $\langle c\rangle$ made by a paper shows a growth
  with the number of published paper $M$ as
  \begin{equation}
  \langle c\rangle = \langle c_0\rangle + b \ln (1+M/M_0),
  \end{equation}
  or
  \begin{equation}
  \langle c\rangle = \langle c_0\rangle + bM/M_0,
  \end{equation}

where $M_0$ is the number of papers published before a reference year and
$\langle c_0\rangle$ is the  average number of citation of that year.
The out-degree distribution of papers, $P_c(c)$   showed that
it had a peak and an exponential decay. While simulating the citation
network this is in fact  the distribution one must use instead  of a power-law distribution
discussed earlier.

In order to qualitatively match with the observed out-degree distribution in \cite{citation}
 we have used a distribution
 \begin{equation}
 P_c(c) \sim  c^{\alpha}\exp(-\beta c)
 \end{equation}
 which has an exponential decay and  a peak at $c_p=\alpha/\beta$.
 Here $c$  can be rescaled to obtain
 \begin{equation}
 P_c(x) \sim x^{\alpha}\exp(-\alpha x)
 \end{equation}
 where $x=c/c_p$.
 The distribution  is then precisely in the form   as in  \cite{citation}
 where the observed value of $\alpha$  is $O(1)$.
 The average number of citations is $\langle c \rangle=(\alpha +1)/\beta$
 and in our simulation we keep $\alpha $ constant and
 vary $\langle c \rangle $ according to eq (5) or (6) such that effectively $\beta$ is  varied
 as time progresses.
 Taking $\alpha = 1$, we generate a network of 4000 papers and choose  $\langle c_0\rangle =5$
 when $M_0$ number of papers have been published
 (this may be considered small, but as the network has accelerated
 growth, the average outdegree increases; we have also checked that the results  do not change
 for other values of   $\langle c_0\rangle$).

 \begin{center}
 \begin{figure}
 \noindent \includegraphics[clip, width=8cm, angle=270]{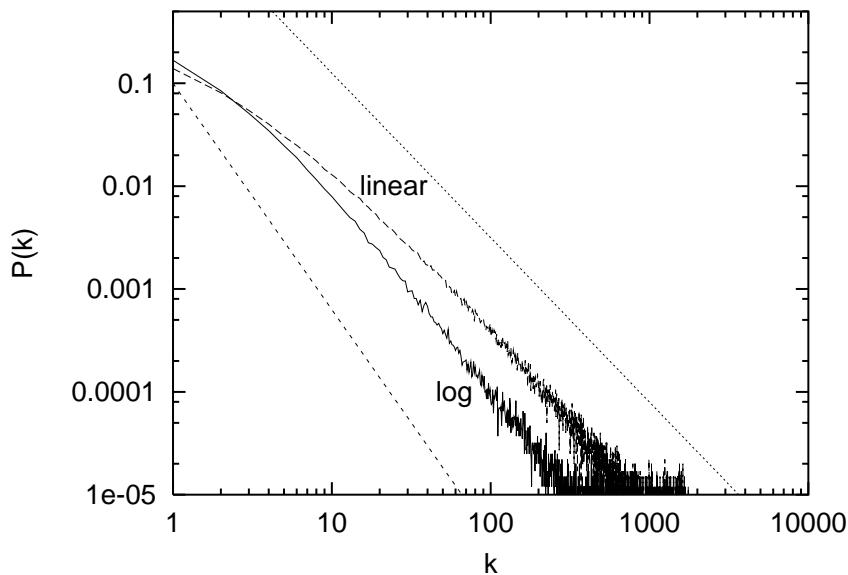}
 \caption{ In-degree distribution $P(k)$ of citation network for linear and logarithmically accelerated citation models.
 The dotted lines have  slope = 2.2  (lower, corresponding to the non-accelerated model) and 1.6 (upper).}
 \end{figure}
 \end{center}

	 The model works in the following way:
	 at any time we have a new paper with $c$ outbound links
	 ($c$ is chosen according to  the probability distribution (eq. 7))
	 which are attached to the existing nodes using a  preferential
	 attachment scheme (eq. 2)
	 with $A=1$.

	 For the logarithmic growth (eq. 5) we find that the degree distribution is
	 scale-free with an exponent close to 2 indicating that the acceleration does
	 not play an important role here.
	 We have used different  values of $b$ (2 and 3) and  $M_0$ (100 and 1000)(comparable to
	 the observed data)   but the results do not
	 depend seriously on these choices.

	 For the linear increase of citations (eq (6)):
	  we find that the behaviour of $P(k)$ does become diffferent
	  from that of the logarithmic case, and the distribution has a power
	  law tail with an exponent $\sim 1.6$.

	   We have also we studied the
	   non-accelerated network ($b=0$). We again find that $P(k)$ has a power law decay with an  exponent close
	   to 2.
	   Hence the accelerated growth does not play a very significant role as far as the logarithmic
	   growth is concerned which is  not very surprising. The linear growth also does
	   not indicate a serious departure from the non-accelerated case possibly because of
	   the low value of the factor $b/M_0$ which controls the effect of the growth term.
	   
	   In the non-accelerated model, the outdegree is a variable but its average is fixed
	   and therefore it is still expected to be described by the general model of \cite{doro2} (this is what
	   happens in the case of the BA model \cite{ba} as observed in \cite{liu}).
	   According to \cite{doro2}  the exponent of $P(k)$ should be $2 + A/\langle c\rangle = 2.2$.
	   We  obtained a
	   slightly smaller value of the exponent;  the disagreement  may be due to the finite size of the system.
	   In fact, the  exponents for the non-accelerated and the logarithmic growth data 
	   are found to be very close.

	   As mentioned earlier, we have not attempted to compare the results with
	   the available citation data.
	   Actually to do so,
	    one should also consider    other factors
	    (e.g., aging \cite{KS},  the fact that  number of papers in each year
	    increases,   increased probability of citing a paper if the contents are closer etc.) in the model.
	    In \cite{citation}, a different algorithm to grow the network had been
	    adopted (the form of the outdegree distribution was not assumed but generated here)
	    which  predicted a power law decay of $P(k)$ with exponent 2.

	    In summary, we have studied
	    two directed networks in which accelerated growth takes place.
	    The number of links is determined by a power-law
	    distribution in the first case and it is more of an academic interest to study it.
	    In the second, we examine the role of accelerated growth in a real
	    network where the distribution of the outgoing link number is chosen according
	    to observed data.
	    The results of the directed model, where $\lambda$ appears as a tuning parameter,
	    are qualitatively similar to the undirected case with the
	    behaviour of the degree distributions showing marked 
	    differnce at different intervals of the value of $\lambda$. 
	    The new
	    feature
	    in the directed accelerated model 
	    is that the degree distributions become  independent of the 
	      value of the initial attractiveness parameter $A$ ($A \neq 0$)
	      as  acceleration is enhanced.

	    It appears that appreciable difference from the 
	    non-accelerated case arises when the out-degree is
	    power-law distributed. 
	   However,
	    such a distribution of out-degree is rare, in real world
	    networks one usually gets an exponential distribution for the
	    out-degree.
	    Also,  in the citation network 
	    the acceleration  term happens to be less significant
	    due to its small magnitude such that the effect of acceleration
	    becomes marginal.

Acknowledgments: 
The author acknowledges the hospitality of ICTP, Trieste,
where part of the work was done. 
Financial support form DST grant SP/S2/M-11/99 is acknowledged.

\end{document}